\documentclass[]{spie}  

\usepackage{amsmath,amsfonts,amssymb}
\usepackage{graphicx}
\usepackage[colorlinks=true, allcolors=blue]{hyperref}
\usepackage{float}
\usepackage{subfigure}

\title{Experimental tests of the calibration of high precision differential astrometry for exoplanets}

\author[a,b,c]{Manon Lizzana}
\author[a]{Fabien Malbet}
\author[a]{Fabrice Pancher}
\author[a]{Sébastien Soler}
\author[d]{Alain Leger}
\author[e]{Thierry Lepine}
\author[a]{Pierre Kern}

\affil[a]{Univ. Grenoble Alpes, CNRS, IPAG, 38000 Grenoble, France}
\affil[b]{Univ. Grenoble Alpes, CNRS, CNES, IPAG, 38000 Grenoble, France}
\affil[c]{Pyxalis (Moirans, France)}
\affil[d]{Univ. Paris-Saclay, CNRS, Institut d’astrophysique spatiale, Orsay, France}
\affil[e]{Institut d’Optique \& Hubert Curien Lab, Univ. de Lyon, Saint-Etienne, France}

\authorinfo{Further author information: \\
E-mail: manon.lizzana@univ-grenoble-alpes.fr}

\begin{document} 
\maketitle

\begin{abstract}
    Current scientific topics like the detection of rocky planets in habitable zone or the study of the dark matter distribution in the Milky Way require differential astrometry measurements with sub-micro arcsecond precision. To achieve this accuracy, the detector and the optical distortion must be calibrated. This paper describes the procedures and the laboratory testbeds developed to carry out the calibrations and then to prepare next space missions.
\end{abstract}

\section{Introduction}
\label{sec:intro}

Astrometry is one of the oldest branches of astronomy which measures the position, the proper motion and the parallax of celestial objects. Differential astrometry allows to increase the precision on a pointed object, the position and motion of the target object are evaluated relatively to the stars present in the field of view. That enables the investigation of new scientific questions like the presence of rocky planets in habitable zone of stars in the Sun vicinity or the study of the nature of dark matter (DM) in the galactic environment \cite{2021ExA....51..845M}. This scientific objectives require sub-micro arcsecond precision, wich can only be achieved with differential astrometry. This is for these reasons that the "Theia" mission was imagined. It has been submitted in 2022 for ESA's M7 call for missions \cite{2022SPIE12180E..1FM}, it is a diffraction-limited telescope about 0.8~m in diameter and with a field of view of 0.5 degrees.\\

The technology readiness level of Theia must be improved manly because the stability requirement on the telescope structure was extremely high and detectors very large. This article will focus on 2 specific aspects : the calibration of new detectors with very large number of pixels, described in Sec. \ref{charac} and \ref{interfero} , and the calibration of the field distortion thanks to the stars in the field of view, described in Sec. \ref{disto}.

\section{Characterize an adequate CMOS detector}
\label{charac}

The scientific goals described below require very large detectors (about 1Gpx) with a low noise level and high sensitivity. For now we are working with Pyxalis, wich is manufacturing such detectors. The aim is to use them for a laboratory demonstration, to ensure that the performance achieved meets the required specifications.\\

\begin{figure}[ht]
    \centering
    \includegraphics[scale=0.55]{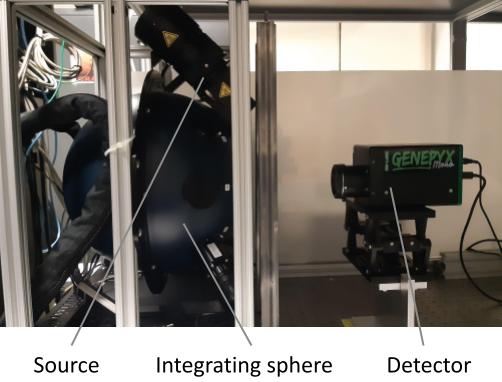}
    \caption{Testbed for the detector characterization}
    \label{fig:benche1}
\end{figure}

The detector used is a 46,000 Gigapyx which is a CMOS with $46$ million pixels of $4.4~\mu\mathrm{m}$. The testbed is displayed on the left of Fig.~\ref{fig:benche1}, the detector is placed in front of an integrating sphere which provides a flat field. The acquisitions have been made at the room temperature.\\

\begin{figure}[ht]
    \centering
    \includegraphics[scale=0.65]{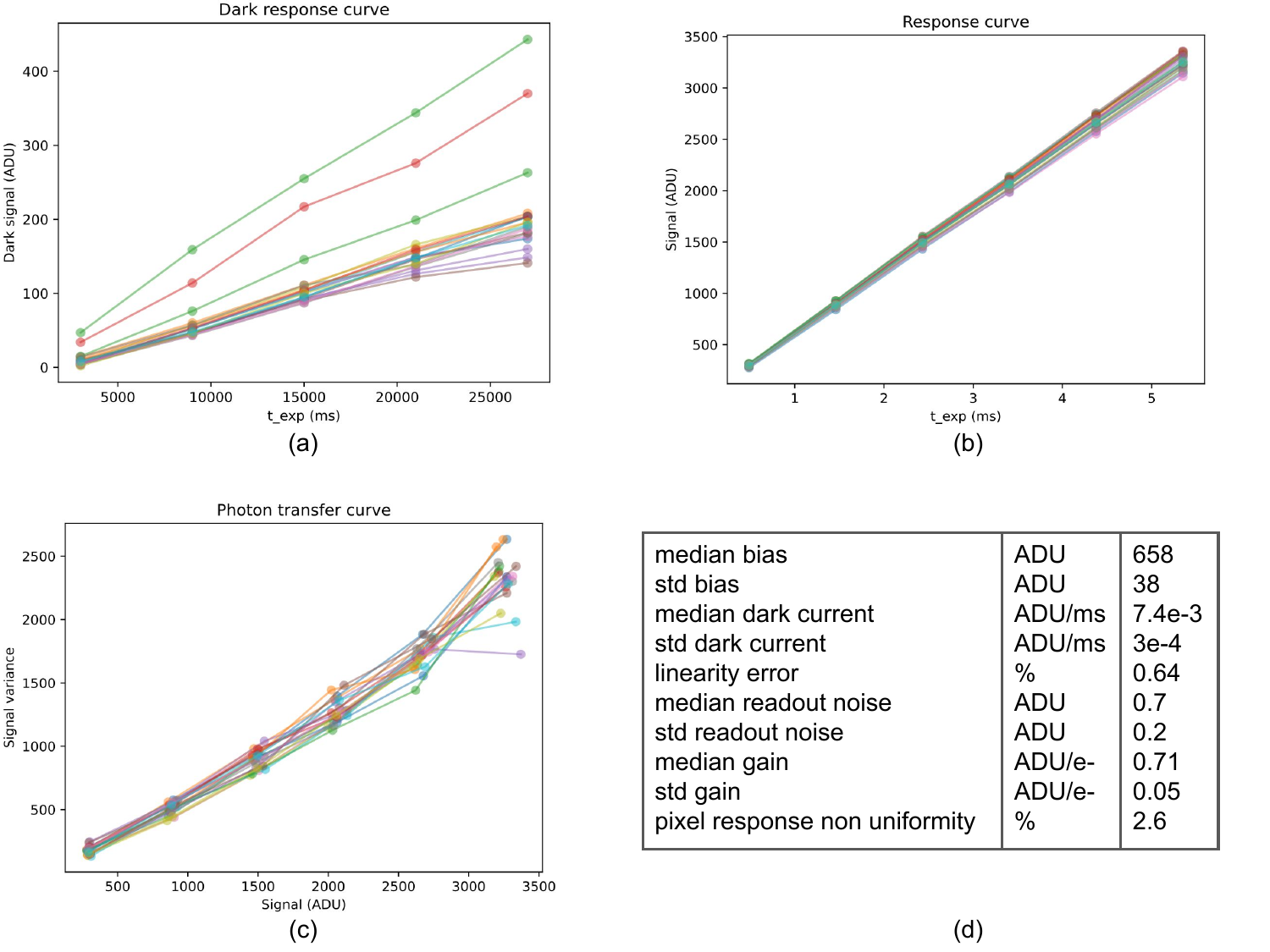}
    \caption{The dark response curve (a), the response curve (b) and the photon transfer curve (c) of the 4600 Gigapyx detector. Only 20 pixels taken randomly have been displayed, each color represent a pixel. The table (d) summarize the detector characterization}
    \label{fig:res_carac1}
\end{figure}

The first results of the global characterization are plotted in Fig.~\ref{fig:res_carac1} and the main figures are summarized. The figures are defined following the standard for characterization of image, sensors and cameras EMVA1288 \cite{EMVA}.\\

First, the bias signal $\mu_{\mathrm{bias\ }i}$ of the frame $i$ is obtained with a negligible exposure time (1~ms) in the dark. These are then subtracted in pairs to give the readout and quantization noise ($\sigma_{\mathrm {roq}}$) :

\begin{equation}
    \sigma=\mathrm{standard\_deviation}(\mu_{\mathrm{bias\ i}}-\mu_{\mathrm{bias\ i+1}})=\sqrt{2} \sigma_{\mathrm {roq}}
    \label{eq1}
\end{equation}

The dark signal versus exposure time is a linear curve and the slope gives the dark current (in ADU/s or $e^-$/s). The same curve, but with the flat signal instead of the dark one, is also linear before saturation (about $5000~e^-$ in high gain and $50000~e^-$ in low gain) and allows the linearity error (LE) of the detector to be calculated. See Eq.(\ref{eq2}), where $n$ is the number of measurements, $\mu_{i}$ the measured flat signal, $t_i$ the exposure time of the acquisition, $a$ and $b$ the parameters of the linear model.

\begin{equation}
    \textrm{LE}=100\frac{1}{n}\sum_{i=1}^{n}\frac{\mid \mu_i-(a t_i+b)\mid}{a t_i+b}
    \label{eq2}
\end{equation}

Furthermore, the slope of the photon transfer curve (variance $\sigma^2$ vs. signal $\mu$) gives the gain/K-factor ($K$) of each pixel (in ADU/$e^-$) : $\sigma^2=K\mu$. Finally, the pixel response non uniformity is calculated by subtracting the noises ($\sigma$) and the signals ($\mu$) with no illumination and at 50\% of the saturation :

\begin{equation}
    PRNU=100\frac{\sqrt{\sigma_{50\%}^2-\sigma_{\mathrm{dark}}^2}}{\mu_{50\%}-\mu_{\mathrm{dark}}}
    \label{eq3}
\end{equation}

\section{Interferometric calibration of the pixel centroid position}
\label{interfero}

\begin{figure}[ht]
    \centering
    \includegraphics[scale=0.5]{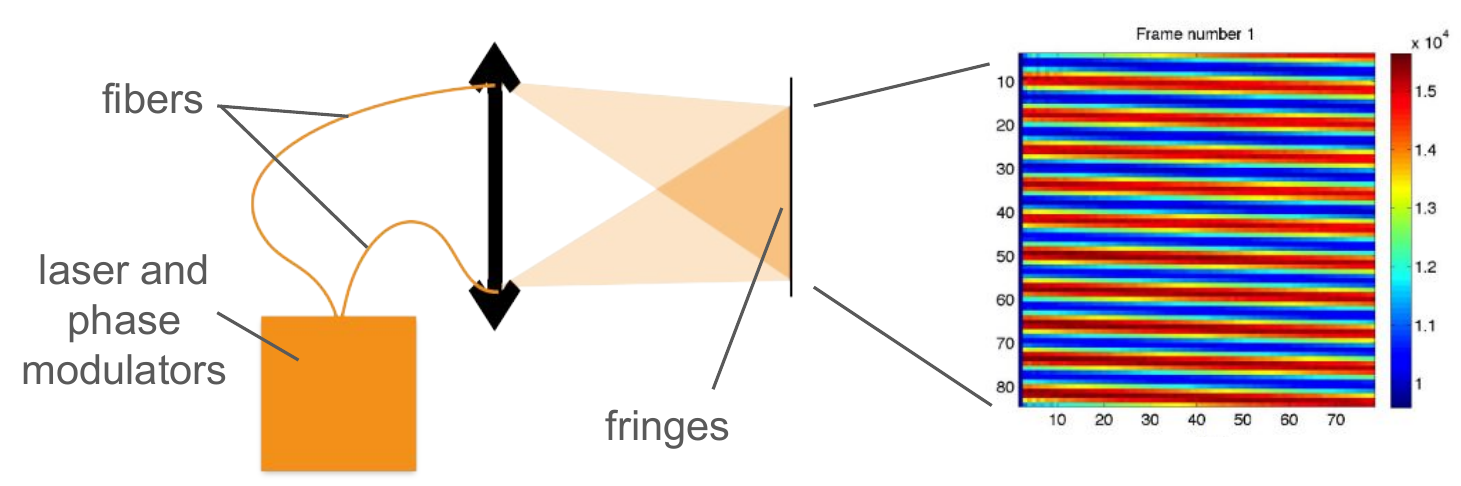}
    \caption{Principle of the interferometric calibration of the pixel centroids position : 2 fibers create interference fringes, the phase modulator makes them scroll, and the detector records the modulation of the flux in each pixel}
    \label{fig:interfero_calib}
\end{figure}

\begin{figure}[ht]
    \centering
    \includegraphics[scale=0.55]{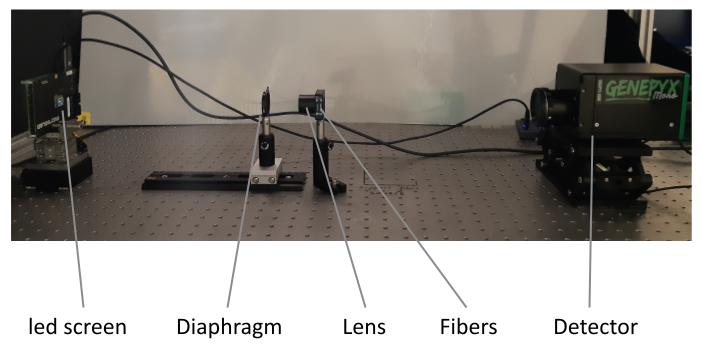}
    \caption{Optical bench for the calibration (see Pancher et al. in this volume for details\cite{poster_fab})}
    \label{fig:benche2}
\end{figure}

To achieve sub-micro arcsecond precision the focal plane must be calibrated spatially with an extreme precision down to the $5.10^{-6}$ pixel level to avoid systematics. In a theoretical detector, rows and columns are well aligned, but in a actual detector the centroids are misaligned (because of fine pixel structure, quantum efficiency local variations ...). Calibrating the detector means evaluating the actual positions of the centroids. As described in Fig.~\ref{fig:interfero_calib}, the general idea is to image Young fringes and make them scroll along the detector. The modulation observed by each pixel provides the positions of the centroids \cite{2023PASP..135g4502S}. Previous work has shown that this is possible with small detector matrices ($80 \times 80$~px) \cite{2016A&A...595A.108C}, the goal is now to check the performances and validate this method with the new very large detectors. The testbed is described in Fig.~\ref{fig:benche2}.

\section{Calibration of optical distortion}
\label{disto}
The optical components of the telescope induce distortions (field distortion, aberrations, drifts from the nominal telescope) that  shift the positions of stars on the detector by thousand of pixels \cite{2022SPIE12180E..1FM}. The required precision is sub-micro-arcsecond corresponding to about $5.10^{-6}$~px on the detector and the stars displacements on the detector due to distortion is larger than $5.10^{-6}$~px, hence the distortion must be calibrated.\\

\begin{figure}[ht]
    \centering
    \includegraphics[scale=0.25]{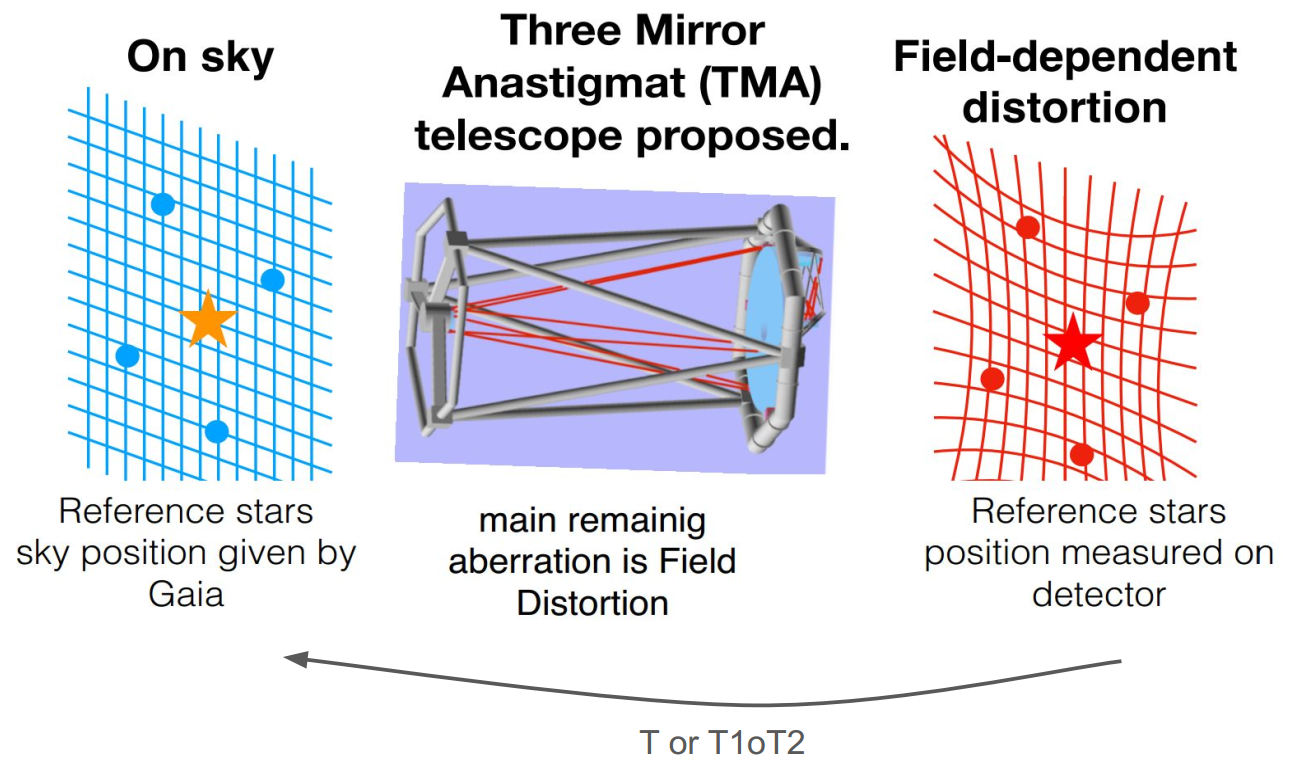}
    \caption{General illustration of the calibration of the distortion. The reference stars are used to adjust a bivariate polynom T that relates the positions of the detector plane to the sky positions. The motion of the target is observed using this transformation.}
    \label{fig:idea_disto}
\end{figure}

Recent work on telescope stability \cite{2022SPIE12180E..1FM} has shown that the reference stars in the field of the telescope can be used as actual metrology sources in order to compute the distortion function of the field. These stars are called 'reference stars' and their astrometric positions are known thanks to the Gaia catalog. The distortion function links the positions of stars on the sky with the positions of star images on the detector, see Fig.~\ref{fig:idea_disto}. This function is modeled by bivariate polynomials.\\

\begin{figure}[ht]
    \centering
    \includegraphics[scale=0.55]{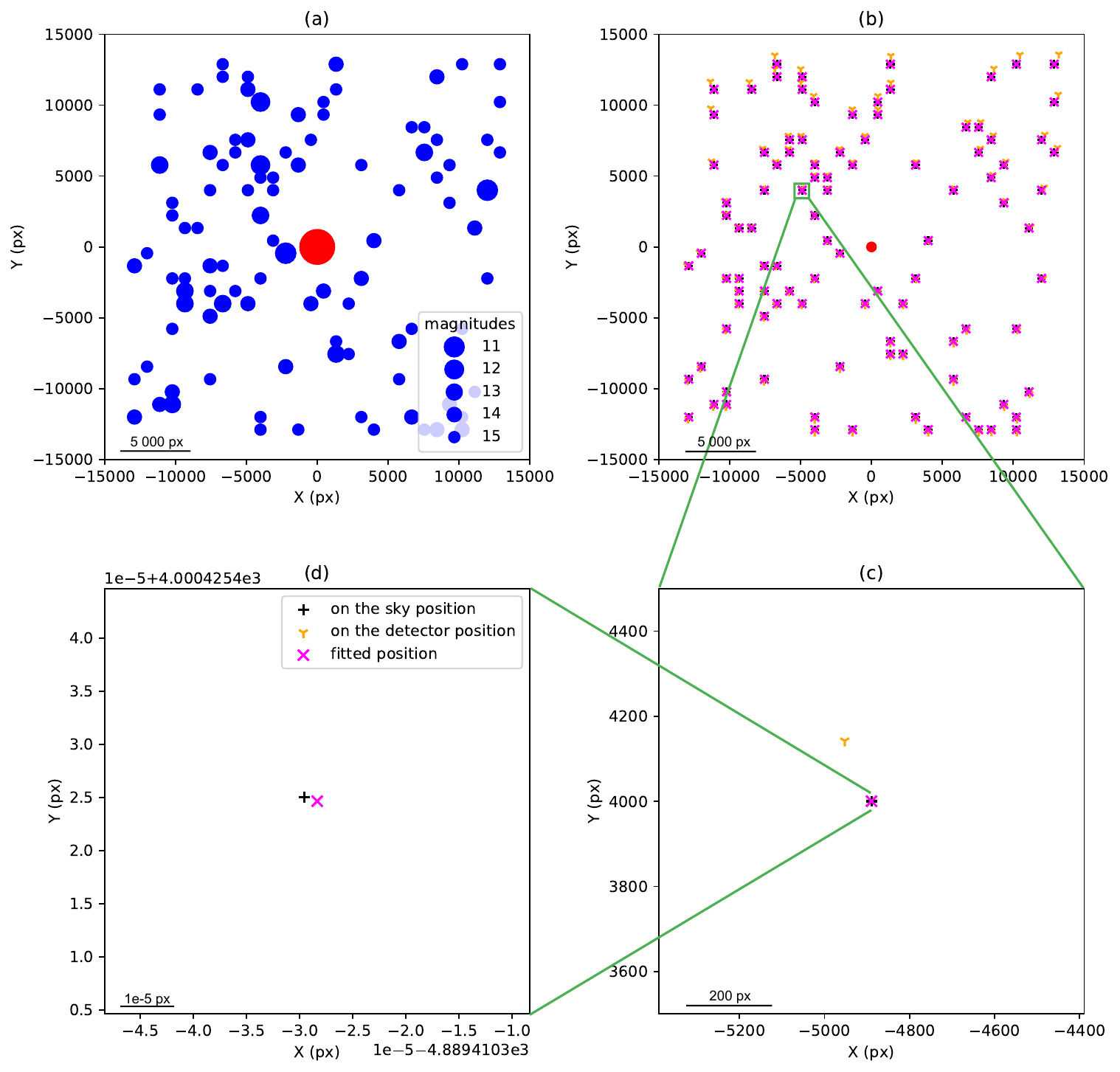}
    \caption{Simulated star field : (a) The image of a star is a circle with size function of its magnitude; the red star is the target star; (b) On-sky and detector positions at full scale : detector positions are slightly shifted compared to on-sky positions due to distortion, but on-sky and fitted positions are indistinguishable; (c) at the scale of the distortion which reach a few hundred pixels, the on-sky and fitted positions are still indistinguishable; (d) at the micropixel scale the on-sky and fitted positions are separated.}
    \label{fig:champ_simu}
\end{figure}

\begin{figure}[ht]
    \centering
    \includegraphics[scale=0.9]{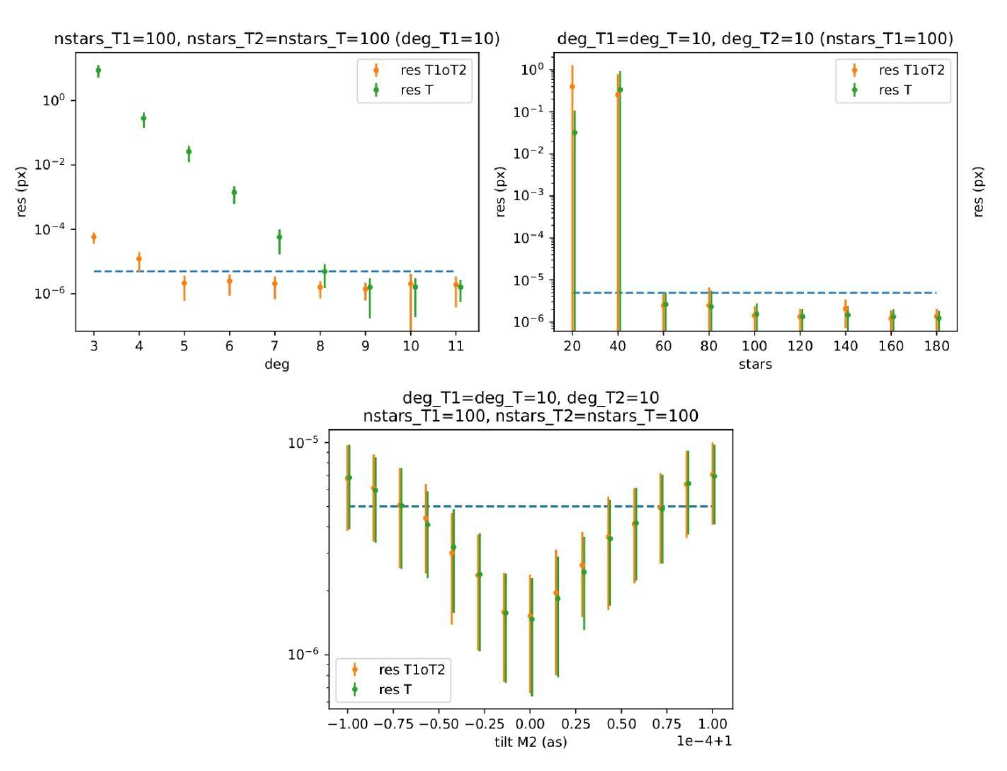}
    \caption{First results of the simulation. Evolution of the residuals with respect to the degree of the polynomial, the number of reference stars and the tilt of the M2.}
    \label{fig:res_simu}
\end{figure}

The positions of the reference stars on the sky $(x_{\mathrm{sky}},y_{\mathrm{sky}})$ are known thanks to the Gaia catalog and the positions on the detector $(x_{\mathrm{det}},y_{\mathrm{det}})$ will be measured by the detector of the telescope as the photocenters of the diffraction spots. The distortion function called $\mathcal{T}(x,y)$ is calibrated with the reference stars. The goal is to precisely measure the movement of the target star without the distortion effects, so the $\mathcal{T}(x,y)$ transformation is applied to the measured position of the target star.\\

A star field is simulated by randomly picking positions on a regular grid $30\times30$ and assigning them an apparent magnitude with a realistic probability calculated with respect the Gaia catalog EDR3 \cite{2021A&A...649A...6G}. The positions on the detector are simulated by optical ray tracing. The polynomial function is then fitted by minimizing the residuals which are the difference between the position on the sky and the position on the detector for each star. Each star is weighted proportionally to the square root of its flux such that $w \propto \sqrt{10^{-m/2.5}}$ with $m$ the magnitude of the star. An example of star field is shown in Fig.~\ref{fig:champ_simu}.\\

In our simulations 3 main parameters are studied : the number of reference stars, the degree of the polynomial, the tilt of the $M_2$ during the exposure time. The main results are presented in Fig.~\ref{fig:res_simu} and show that the stars in the field of view can be used as metrology sources. In our simulations a precision of $5.10^{-6}$~px can be reached with about 100 stars and an $8^{\text{th}}$ order polynomial, if the M2 tilt is below $7.10^{-5}$~as for that exposure. In addition, the residuals are almost constant all over the field of view.\\

For now some parameters are not taken into account like the Gaia uncertainties, the photon noise or the dark current of the detector. We do not expect the conclusions to vary a lot but that must be checked in future simulations. The simulations have to be improved to provide more robust conclusions.\\

A testbed is currently developed to experimentally show the performances of this new field calibration method. It is described on the right of Fig.~\ref{fig:benche2}. This experiment will enable us to study the performances of the field calibration method and show that the procedure described before works in laboratory. In practice, pseudo stars are illuminated on a led screen and a distortion is induced with a lens and a mooving diaphragm. Distorted star images are recorded with the detector and the previous procedure to recover the undistorted image is applied.

\section{Conclusion}

In this paper, we described the general characterisation of the Gigapyx, a new very large detector of 46~Mpx. Then we presented the interferometric procedure which will be applied in laboratory to calibrate the centroid position of pixels. Finally, as the optical distortion must be calibrated, we described on the one hand a simulation which models the distortion function with a polynomial, and on the other hand the laboratory experiments which will be carry out to measure the accuracy of the procedure.

\acknowledgments

This work has been partially supported by the LabEx FOCUS ANR-11-LABX-0013 and the CNES agency. Manon Lizzana would like to acknowledge the support of her PhD grant from CNES and Pyxalis.

This research has made use of NASA’s Astrophysics Data System Bibliographic Services and of the CDS, Strasbourg Astronomical Observatory, France.

\bibliography{report} 
\bibliographystyle{spiebib} 

\end{document}